\documentclass[11pt]{article}\usepackage{theorem, amsfonts,amsmath,epsfig}
\title{Hydrodynamics in an external field}
\author{M. R. Grasselli and R. F. Streater,\\Dept. of Mathematics,\\
King's College London,\\Strand, London WC2R 2LS.\\}
\date{14/05/2001}
\begin{document}
\maketitle
\begin{abstract}
The methods of statistical dynamics are applied to a fluid
with 5 conserved fields (the mass, the energy, and three
components of momentum) moving in a given external field. When the
field is zero, we recover a previously derived system of parabolic
partial differential equations, called `corrections to fluid dynamics'.
\end{abstract}

\vspace{.1in}
\noindent{\bf Key Words:} Navier-Stokes, bulk diffusion, Soret, Dufour,
thermal diffusion.
\section{Introduction}

In a previous paper \cite{streater4}, we described a stochastic
model of a fluid with no external field and derived a system
of parabolic equations expressing the dynamics of the
density-fields of mass, energy and momentum. An unusual feature of
the system is that the Euler continuity equation acquires a
diffusion term. This bulk diffusion does not appear in the standard theory
\cite{balescu}, but has arisen in some other work \cite{beck,dobrushin,xing}.
It is likely that our equations are more stable
and physically more accurate than the usual Navier-Stokes
equations. In a model in an external field, but without a velocity field
\cite{streater1}, we studied the dynamics of the mass
and energy densities; they
differed from the equations got by putting the velocity field,
${\bf u}({\bf x},t)$ equal to zero in the Navier-Stokes system. In
particular, the mass-density $\rho({\bf x},t)$ did not obey the
Euler continuity equation with ${\bf u}=0$, but rather the Smoluchowski
equation. This contains not only the diffusion term found in
\cite{streater4}, but also a drift, as predicted by Smoluchowski,
and implicit in the work of Einstein. In the present paper, we
take the model of \cite{streater4}, and put it in an external
field $\Phi$. We derive the full set of equations
\begin{eqnarray}
\frac{\partial\rho}{\partial t} &+& \mbox{ div}({\bf u}\rho)
= \lambda\mbox{ div}\left(\rho^{-1}\nabla (\Theta^{1/2}\rho) +
\frac{\nabla\Phi}{k_B\Theta^{1/2}}\right)\\
\frac{\partial(\rho e)}{\partial t} &+&  \mbox{ div }[{\bf u}(\rho e +
P)] = \lambda\mbox{ div }\left[2\rho^{-1}\nabla\left(\Theta^{1/2}P\right)
+ \rho^{-1}\nabla\left(\Theta^{1/2}\rho\right)\Phi/m \right.\nonumber \\
&+& \left.\frac{\nabla\Phi}{k_B\Theta^{1/2}}\phi + 2\frac{\nabla\Phi}
{k_B\Theta^{1/2}}P \right] \\
\frac{\partial\rho{\bf u}}{\partial t}&+&\mbox{div}(\rho{\bf u}\otimes{\bf u})
=-\rho\nabla \Phi/m-\nabla P +\lambda\mbox{ div}\left(\frac{\nabla\Phi}
{k_B\Theta^{1/2}}\otimes{\bf u}\right) \nonumber \\
&+&\frac{2\lambda}{5}\partial^i\rho^{-1}\left[3\partial_i\left(\rho(x)
\Theta^{1/2}{\bf u}\right)
+\nabla\left(\rho(x)\Theta^{1/2}u_i\right)\right].
\end{eqnarray}
Here, $\rho$ is the mass density, ${\bf u}$ is the velocity field,
and $\Theta$ is the temperature. The energy density per unit mass,
$e$, reduces in this approximation to $\Phi/m+3k_B\Theta/(2m)$.
The equations  reduce to
those of \cite{streater4} when $\Phi=0$, while if ${\bf u}=0$ they
reduce to \cite{streater1} with suitable changes due to the
different multiplicity of states in the two models. This is a
property of any good approximation method, which is not true
of the usual compressible Navier-Stokes equations with temperature,
as studied in \cite{lions1,lions2} for example.
The equations of \cite{streater1} can be extended to models with
interparticle forces, treated as a nonlinear mean field theory by
a development of the theory of non-linear parabolic systems
\cite{biler1,biler2,biler3,biler4,biler5}.

\section{Review of the model}

\subsection{The sample space, conserved quantities and information manifold}

Let $\Lambda$ be a finite subset of the cubic lattice
$(a\mathbb{Z})^3$ with spacing $a \approx 10^{-8}$ cm,
representing the size of the hard core of the fluid molecules. We
endow our model with a local structure by assigning to each $x \in
\Lambda$ a sample space $\Omega_x$, which specifies the possible
configurations at $x$. The total sample space is then the product
space $\Omega=\prod_{x\in\Lambda} \Omega_x$. For a structureless
monatomic fluid (e.g. argon) we choose
\begin{equation}
\Omega_x=\left\{\emptyset,(\epsilon\mathbb{Z})^3\right\},
\end{equation}
where $\epsilon \approx 6.6\times 10^{-19}$ c.g.s. is the quantum
of momentum of a particle confined to a region of size $a^3$ (see
the discussion following equation (6) in \cite{streater4}) . A
point $\omega\in\Omega$ is thus the collection
$\{\omega_x\}_{x\in\Lambda}$. If $\omega$ is such that
$\omega_x=\emptyset$ for a certain $x\in\Lambda$ then the
configuration $\omega$ has no particle at $x$; we also say that
there is a hole, or vacancy, at $x$. If $\omega_x= {\bf
k}\in(\epsilon\mathbb{Z})^3$, then in the configuration $\omega$
there is a particle at $x$ and its momentum is ${\bf k}$.

The state space of the system consists of the probability measures
on $\Omega$ and is denoted by $\Sigma(\Omega)$. The {\em
information manifold} ${\cal S}\subset\Sigma$ associated with the
model consists of the parametric exponential family
\cite{grasselli1} determined by the slow variables, that is, the
local mean values of the five conserved quantities: the mass, energy
and the three components of the momentum along the three unit
vectors $e_1,e_2,e_3$ generating the lattice. These are the random
variables
\begin{eqnarray}
{\cal N}_x(\omega)&=&\left\{\begin{array}{ll}
0&\mbox{ if $\omega_x=\emptyset$}\\
1&\mbox{ if $\omega_x={\bf k}\in(\epsilon\mathbb{Z})^3$}
\end{array}\right.\\
{\cal E}_x(\omega)&=&\left\{\begin{array}{ll}
0&\mbox{ if $\omega_x=\emptyset$}\\
(2m)^{-1}{\bf k.k}+\Phi(x)&\mbox{ if $\omega_x={\bf k}$}
\end{array}\right.\\
\mbox{\boldmath${\cal P}$}_x(\omega)&=&\left\{\begin{array}{ll}
0&\mbox{ if $\omega_x=\emptyset$}\\
{\bf k}&\mbox{ if $\omega_x={\bf k}$},
\end{array}\right.
\end{eqnarray}
where $m$ is the mass of the molecule and $\Phi$ is a given real-valued
potential, which could be time-dependent. The states in ${\cal S}\subset\Sigma$ are those
of the form $p=\prod_x p_x$, where
\begin{equation}
p_x=\Xi_x^{-1}\exp\left\{-\xi_x{\cal N}_x-\beta_x{\cal E}_x-
\mbox{\boldmath$\zeta$}_x\cdot\mbox{\boldmath${\cal
P}$}_x\right\}.
\end{equation}
Here, $\xi_x,\beta_x$ and $\mbox{\boldmath$\zeta$}_x$
are fields of intensive variables that obviously determine $p$;
they are called the canonical coordinates for $p\in {\cal S}$. The
great grand partition function $\Xi_x$ at each site is the
normalising factor
\begin{eqnarray}
\Xi_x &=& \sum_{\omega_x\in\Omega_x}\exp\left\{-\xi_x {\cal N}_x-\beta_x{\cal
E}_x-\mbox{\boldmath$\zeta$}_x \cdot\mbox{\boldmath${\cal P}$}_x\right\} \\
 &=& 1+e^{-\xi_x-\beta_x\Phi(x)}Z_1Z_2Z_3,
\end{eqnarray}
where
\begin{equation}
Z_i=\sum_{k\in\epsilon\mathbb{Z}}\exp\left(-\frac{\beta_x
k^2}{2m}-\zeta_x^i k\right).
\end{equation}
This can be calculated explicitly if, due to the small parameter
$\epsilon$, we approximate sums by integrals:
\begin{eqnarray}
Z_i &\approx& \epsilon^{-1}\int_{-\infty}^{\infty} e^{-\frac{\beta_x
k^2}{2m}-\zeta_x^i k}dk \\
&=&
\epsilon^{-1}\left(\frac{2m\pi}{\beta}\right)^{1/2}e^{m(\zeta^i)^2/(2\beta)}.
\end{eqnarray}

The means $\left(N_x,E_x,\mbox{\boldmath$\varpi$}_x\right)$ of the slow variables in a
state $p\in M$ also determine $p$; these are called the `mixture'
coordinates. They are related to the canonical coordinates by a
Legendre transform:
\begin{eqnarray}
N_x&=&-\frac{\partial\log\Xi_x}{\partial\xi_x},\hspace{.6in}x\in\Lambda\\
E_x&=&-\frac{\partial\log\Xi_x}{\partial\beta_x},\hspace{.6in}x\in\Lambda\\
\varpi_x^i&=&-\frac{\partial\log\Xi_x}{\partial\zeta_x^i},
\hspace{.6in}i=1,2,3,\;\;x\in\Lambda.
\end{eqnarray}

Using the explicit expression for the partition function, these
can be used to deduce several equations relating the macroscopic
variables in the theory. In particular, if we introduce the mean
velocity field
\begin{equation}
\mathbf{u}_x=\frac{\mbox{\boldmath$\varpi$}_x}{mN_x},
\end{equation}
and the temperature field $\Theta_x=1/(k_B\beta_x)$, it is
straightforward to show that
\begin{equation}
\zeta^i=-\frac{\beta_x\varpi^i_x}{mN_x}=-\beta_x u^i_x
\end{equation}
and
\begin{equation}
E_x=N_x\left(\Phi(x)+\frac{3}{2}k_B\Theta_x +
\frac{1}{2}m\mathbf{u}_x\cdot\mathbf{u}_x \right).
\label{Nenergy}
\end{equation}

The discrete nature of the model allows us to use the von Neumann
entropy
\begin{equation}
S(p):=-k_B\sum_\omega p(\omega)\log p(\omega),
\end{equation}
which for $p\in {\cal S}$ gives
\begin{equation}
S=\sum_{{\bf x}\in\Lambda}\left(\xi_{\bf x}N_{\bf x}+\beta_{\bf
x}E_{\bf x}
+\mbox{\boldmath$\zeta_x$}\cdot\mbox{\boldmath$\varpi_x$}+
\log\Xi_{\bf x}\right)
\end{equation}
An argument from equilibrium theory \cite{streater4} then leads to
the definition of the thermodynamical pressure as
\begin{equation}
P({\bf x})=a^{-3}k_B\Theta\log\Xi_{\bf x}.
\end{equation}
If the ratio $V_0/V$ (between the smallest volume that the
$N=\sum_x N_x$ particles can occupy, that is $V_0=a^3N$, and the
total volume $V$) is small, then the formula above for the pressure
reduces to the perfect gas approximation
\begin{equation}
P_x=\frac{N_xk_B\Theta_x}{a^3}
\end{equation}

The other macroscopic variables in terms of which the
hydrodynamical equations are written are the mass density
$\rho(x)=mN_x/a^3$ and the energy-density per unit of mass
$e(x)=E_x/(mN_x)$. If we ignore the small term involving
${\bf u}_x\cdot {\bf u}_x$ in (\ref{Nenergy}), we see that
\begin{equation}
e(x)=\frac{\Phi(x)}{m}+\frac{3k_B\Theta_x}{2m}
:=\phi(x)+\frac{3k_B\Theta_x}{2m}.
\end{equation}

\subsection{The hopping dynamics and the continuum limit}

Whether time is discrete or continuous, the dynamics traces out an
orbit in ${\cal S}$. The need to take the time-interval larger
than zero has been well explained in \cite{balian}.
For discrete time, the dynamics consists of
two steps. The first one is a stochastic map $T:\Sigma\rightarrow
\Sigma$ which maps each shell
\[\Omega_{N,E,\mbox{\boldmath$\varpi$}}=\left\{\omega\in\Omega :
\sum_x {\cal N}_x=N,\sum_x {\cal E}_x=E,\sum_x
\mbox{\boldmath${\cal P}$}_x=\mbox{\boldmath$\varpi$}\right\}\] to
itself. This is obtained when we specify the hopping rules that
are responsible for changing the configuration of the particles
and hence the local values of the slow variables. If we start with
a state $p\in {\cal S}$, the state $T(p)$ will generally not be an
exponential state. The second step in the dynamics is then a
orthogonal projection back to ${\cal S}$ following a path that
conserves the means of all the slow variables. This is the
thermalising map $Q$, defined for any state $p$ with finite
expectations of the five slow fields, as the unique point in
${\cal S}$ with these expectations as mixture coordinates. So $Qp$
is determined by
\begin{eqnarray}
{\bf E}_{Qp}[{\cal N}_x]&=&{\bf E}_p[{\cal N}_x]=N_x\\
{\bf E}_{Qp}[{\cal E}_x]&=&{\bf E}_p[{\cal E}_x]=E_x\\
{\bf E}_{Qp}[\mbox{\boldmath${\cal P}$}_x]&=&{\bf
E}_p[\mbox{\boldmath ${\cal P}$}_x]=\mbox{\boldmath$\varpi$}_x.
\end{eqnarray}
The precise meaning of orthogonality in the description above, as
well as the characterisation of the path followed during the
projection as a geodesic for a certain affine connection (the
`mixture' connection), is part of the subject called {\em
information geometry}
\cite{balian,ingarden1,ingarden2,jaynes,kossakowski,grasselli1,grasselli2}.

Our model is specified by giving hopping rules. We require that
$T$ should couple only neighbouring points in $\Lambda$, where we
consider two points to be neighbours if their distance along one
of the lattice unit vectors is one mean free path, denoted by
$\ell$. A particle moving from a site is taken to move exactly the distance
$\ell$, and then to thermalise. This was called {\em abrupt thermalisation}.
The more elaborate assumption, that the size
of its hop is random, governed by the exponential law with mean $\ell$,
leads \cite{RFS7} to similar conclusions. We assume that $\ell$ is an
integer multiple of the lattice spacing $a$, but allow it to depend on the
local density by taking $\ell/a$ to be the nearest integer to
\[\frac{\rho_{max}}{\rho(x)}=\frac{m}{a^3\rho(x)}.\]
Suppose that $\omega \in \Omega$ is such that $x\in\Lambda$ is
occupied. Consider in turn the possibility of jumping from
$x$ along the direction of the unit vectors $\pm e_i$ of the cubic
lattice to an empty site $x^\prime:=x\pm\ell e_i, i=1,2,3$.
In the absence of an external potential \cite{streater4},
the jump will take a time
$\ell/|v^i_x|$, where ${\bf v}_x:={\bf k}_x/m$. We then define the (random)
hopping {\em rate} from $x$ to $x+\ell e_i$ to be
the inverse of this relaxation time, namely $v^i_x/\ell$ if $v^i_x\geq 0$ and
zero if $v^i_x$ is negative, in which case there is a rate $-v^i_x/\ell$ of
hopping to $x-\ell e_i$. The situation in the presence of an
external potential is a little more involved, because the potential causes a
change in the velocity along the jump. From now on, assume for definiteness
that $\Phi(x+\ell e_i)>\Phi(x)$. If the particle at $x$ hops to
$x+\ell e_i$, its potential energy increases to $\Phi(x+\ell e_i)$
and so its kinetic energy must decrease by the same
amount. Its change in momentum is taken to be entirely in the
direction of $e_i$. So its velocity in the $i$-direction,
$v_x:=k_x/m$ (we omit the index $i$ for $v$ and $k$ in the following
formulae since it is clear from the rest of the notation what is the
component involved) is reduced to $v_x^\prime:=k_x^\prime/m$, where
\begin{eqnarray}
k_x^{\prime 2}/(2m) &=& k_x^2/(2m)-\Phi(x+\ell e_i)+\Phi(x) \nonumber \\
&=& k_x^2/(2m)-\ell\partial_i\Phi(x)+O(\ell^2),
\end{eqnarray}
that is
\begin{equation}
k_x^\prime = \left(k_x^2-2m\ell\partial_i\Phi(x)\right)^{1/2}.
\label{kprime}
\end{equation}
In order for the move to be energetically possible, we must have
\begin{equation}
k_x\geq\kappa^i_x:=\left(2\ell m\partial_i\Phi(x)\right)^{1/2}.
\end{equation}
Similarly, if the particle at $x$ with velocity $v_x<0$ in the $i$-direction
hops to $x-\ell e_i$, its potential energy decreases to $\Phi(x-\ell e_i)$,
with a corresponding rise in its kinetic energy.
Therefore (again taking the change in momentum to be entirely
in the $i$-direction) its (negative) velocity in the $i$-direction
becomes $v_x^{\prime\prime}=k_x^{\prime\prime}/m$, where
\begin{eqnarray}
k_x^{\prime\prime 2}/(2m) &=& k_x^2/(2m)+\Phi(x)-\Phi(x-\ell e_i) \nonumber \\
&=& k_x^2/(2m)+\ell\partial_i\Phi(x-\ell e_i)+O(\ell^2),
\end{eqnarray}
that is
\begin{equation}
k_x^{\prime\prime}=-\left(k_x^2+2m\ell\partial_i\Phi(x-\ell e_i)\right)^{1/2}.
\label{kprimeprime}
\end{equation}
We take the hopping rate
from $x$ to $x^\prime=x\pm\ell e_i$ to be the average of the initial and
final rates:
\begin{equation}
r(k_x)=\left\{\begin{array}{cl}
r_-(k_x):=-\frac{v_x+v_x^{\prime\prime}}{2m\ell}=
\frac{-k_x+\left[k_x^2+(\kappa^i_{x-\ell e_i})^2\right]^{1/2}}{2m\ell},
&\mbox{ if }k_x\leq 0;\\
r_+(k_x):=\frac{v_x+v_x^\prime}{2m\ell}=
\frac{k_x+\left[k_x^2-(\kappa^i_x)^2\right]^{1/2}}{2m\ell},
&\mbox{ if }k_x\geq\kappa^i_x.
\end{array}\right.
\label{rates}
\end{equation}
These hopping rates increase
with $k_x$ and there are infinitely many possible momentum states. To be
a Markov chain, the sum of all rates out of a configuration must be
less than one. For any $k_x$ this can be achieved by choosing $dt$ small
enough. To do this for all $k_x$ with a fixed $dt$ we must put in a cut-off;
there are no hops if $|k_x|>K_x$, say. Finally, $r(k_x)dt$ gives the
probability of a transition in an interval $dt$ provided that the site
$x$ is occupied and the site $x^\prime$ is empty, so the actual
entries of the Markov matrix are conditional probabilities and the transition
rate above should appear multiplied by factors of the form
$N_x(1-N_{x^\prime})$.
As argued in \cite{streater4}, we neglect $N_x$ comparing
to 1, therefore leaving out the second term in the factors above.

The continuum limit we are going to take in order to obtain the
hydrodynamical equations corresponds to $\ell\rightarrow 0$, $c\rightarrow
\infty$ such that the product $\ell c$ remains finite and non-zero, where
\begin{equation}
c:=\left(\frac{k_B\Theta_0}{m}\right)^{1/2}
\end{equation}
is the approximate velocity of sound at the reference temperature $\Theta_0$.
The diffusion constant that appears when we take the limit is then predicted
to be
\begin{equation}
\lambda:=\frac{\ell c \rho}{(2\pi\Theta_0)^{1/2}}
= \frac{a\rho_{max} c}{(2\pi\Theta_0)^{1/2}}.
\end{equation}

\section{Hydrodynamics in an external field}

When a transition from $x$ to $x+\ell e_i$ occurs in a potential $\Phi$,
the loss of mass and energy from the site $x$ is equal to the gain at
the site $x+\ell e_i$. This is not true of momentum; the loss at $x$
differs from the gain at $x+\ell e_i$ by $\kappa_i:=(2\ell m\partial_i
\Phi)^{1/2}$. So we deal will ${\cal N}$ and ${\cal E}$ first. We take it
that if $0\leq k^\prime<\kappa_i$ then no hop is made.

Before we start the calculations, let us recall that integrals of the form
\[M_n(\zeta)=\int_0^{\infty}k^n\exp\{-\beta k^2/(2m)-\zeta k\}dk, \qquad
n=0,1,2,3\]
were evaluated up to second order in $\zeta$ in Appendix 1 of
\cite{streater4}.
For later use, we reproduce the results here up to zeroth order in $\zeta$ for
$n=0,1$,
\begin{eqnarray}
M_0(\zeta) &=& \left(\frac{\pi m}{2\beta}\right)^{1/2}, \\
M_1(\zeta) &=& \frac{m}{\beta},
\end{eqnarray}
and to first order in $\zeta$ for $n=2$,
\begin{equation}
M_2(\zeta) = \left(\frac{\pi}{2}\right)^{1/2}
\left(\frac{m}{\beta}\right)^{3/2}-2\left(\frac{m}{\beta}\right)^2\zeta.
\end{equation}

\subsection{Dynamics of the mass-density in an external field}

Since the field $\Phi(x)$ is external, it does not depend on the
configuration $\omega_x$ of the random fields at $x$.
It therefore cancels
in the exponential states. This is seen in its simplest case in the model
studied in \cite{streater}. The potential enters only in its supression or
enhancement of the transition rate;  in the present case, the rate is the
average of the initial and final rates. This shows up mainly in the
appearance of a non-zero lower limit to the (positive) momentum for
any right-going hop to be possible.

Let $J_x^i/\ell$ be the change in the value of $N_x$ due to the
hoppings occuring between $x$ and $x-\ell e_i$ in such a
way that
the change due to exchanges with both $x\pm \ell e_i$
in an interval $\delta t$ is
\begin{equation}
\delta_i N_x = -\frac{J^i_{x+\ell e^i}-J^i_x}{\ell}\delta t.
\label{iNchange}
\end{equation}
So the total change in $N_x$ in an interval $\delta t$ due to hoppings
in all directions is
\begin{equation}
\delta N_x = (\delta_1 N_x + \delta_2 N_x + \delta_3 N_x)\delta t.
\label{Nchange}
\end{equation}

Using the hopping rates defined in the previous section, the loss/gain
contribution to the particle current involving the exchange between
$x$ and $x-\ell e_i$ is
 \begin{equation}
J^i_x=-\sum_{k^i\leq 0} \ell r_-(k_x)p_x({\bf k}){\cal N}_x({\bf k}) +
\sum_{k^i\geq \kappa^i} \ell r_+(k_{x-\ell e_i})p_{x-\ell e_i}(k)
{\cal N}_{x-\ell e_i}(k).
\label{jota}
\end{equation}

As in \cite{streater4}, the analysis of this expression is best handled by
introducing a conditional probability
$\bar{p}_x(\omega)=p(\omega|{\cal N}_x=1)$ on the particle space
$\Omega-\emptyset$, that is,
\begin{equation}
\bar{p}_x({\bf k})=(Z_1Z_2Z_3)^{-1}\exp\{-\beta_x|{\bf k}|^2/(2m)-
\mbox{\boldmath$\zeta$}_x\cdot {\bf k}\}.
\end{equation}

We now use the fact that $p_x(k)=N_x\bar{p}_x(k)$ and ${\cal N}_x(k)=1$ on the
particle space $\Omega-\emptyset$, replace the sums by integrals in
(\ref{jota}), and add and subtract the term
\begin{equation}
F^i_x=N_x(Z_i\epsilon)^{-1}\int_{k\geq\kappa}\ell r_+(k)\exp\left\{-\frac
{\beta k^2}{2m}-\zeta k\right\}dk,
\label{small}
\end{equation}
to obtain
\begin{eqnarray}
J^i_x &=& F^i_x -\frac{N_x}{Z_i\epsilon}\int_{k\leq 0}\ell r_-(k_x)
\exp\left(-\frac{\beta k^2}{2m}-\zeta^i k\right)dk\nonumber\\
&& -\ell\left(F^i_x-F^i_{x-\ell e_i}\right)/\ell.
\label{J}
\end{eqnarray}

We start by calculating $F^i_x$. In the term
\[\int_{k\geq\kappa}\frac{k+(k^2-\kappa^2)^{1/2}}{2m}
\exp\left\{-\frac{\beta k^2}{2m}-\zeta^ik\right\}dk,\]
we make the change of variable
\[k^{\prime 2}=k^2-\kappa^2.\]
Then $k\,dk=k^\prime\,dk^\prime$, and the integral becomes
\[\int_{k^\prime\geq 0}\frac
{k^\prime \left[(k^{\prime 2}+\kappa^2)^{1/2}+k^\prime\right]}
{2m(k^{\prime 2}+\kappa^2)^{1/2}}
\exp\left\{-\frac{\beta (k^{\prime 2}+\kappa^2)}{2m}-\zeta^i
(k^{\prime 2}+\kappa^2)^{1/2}\right\}dk^\prime,\]
which can be written as
\begin{eqnarray*}
\lefteqn{\int_{k^\prime\geq 0}\frac
{k^\prime \left[(k^{\prime 2}+\kappa^2)^{1/2}+k^\prime\right]}
{2m(k^{\prime 2}+\kappa^2)^{1/2}}
\exp\left\{-\frac{\beta k^{\prime 2}}{2m}-\zeta^ik^\prime\right\}}\\
&\times &\exp\left\{-\frac{\beta\kappa^2}{2m}-\left((k^{\prime 2}
+\kappa^2)^{1/2}-k^\prime\right)\zeta^i\right\}dk^\prime.
\end{eqnarray*}
The arguments of the exponentials are small, and we expand them to first order:
\[\exp\left\{-\frac{\beta\kappa^2}{2m}
-\left[\left(k^{\prime 2}+\kappa^2\right)^{1/2}-k^\prime\right]
\zeta^i\right\} = 1-\frac{\beta\kappa^2}{2m}-
\left[k^\prime-\left(k^{\prime
2}+\kappa^2\right)^{1/2}\right]\zeta^i.\] This gives us the three
terms
\begin{eqnarray}
& &\int_{k^\prime\geq 0}
\frac{k^\prime \left[(k^{\prime 2}+\kappa^2)^{1/2}+k^\prime\right]}
{2m(k^{\prime 2}+\kappa^2)^{1/2}}
\exp\left\{-\frac{\beta k^{\prime 2}}{2m}-\zeta^i k^\prime\right\}dk^\prime
\label{A}\\
& &-\frac{\beta\kappa^2}{2m}\int_{k^\prime\geq 0}
\frac{k^\prime \left[(k^{\prime 2}+\kappa^2)^{1/2}+k^\prime\right]}
{2m(k^{\prime 2}+\kappa^2)^{1/2}}
\exp\left\{-\frac{\beta k^{\prime 2}}{2m}-\zeta^ik^\prime\right\}dk^\prime
\label{B}\\
& &-\frac{\kappa^2\zeta^i}{2m}\int_{k^\prime\geq 0}
\frac{k^\prime}{(k^{\prime 2}+\kappa^2)^{1/2}}
\exp\left\{-\frac{\beta k^{\prime 2}}{2m}-\zeta^ik^\prime\right\}dk^\prime.
\label{C}
\end{eqnarray}
The dominant term is (\ref{A}), in which we may replace the factor
\[\frac{k^\prime \left[(k^{\prime 2}+\kappa^2)^{1/2}+k^\prime\right]}
{2m(k^{\prime 2}+\kappa^2)^{1/2}}\]
by the velocity when $\Phi=0$, namely $k^\prime/m$, with an error of
$O(\ell\log\ell)$ (Appendix 1). So the contribution
of this term to the mass current can be approximated in the limit by
\begin{equation}
\frac{N_x}{Z_i\epsilon}\int_0^{\infty}\frac{k^\prime}{m}
\exp\left\{-\frac{\beta k^{\prime 2}}{2m}-\zeta^ik^\prime\right\}dk.
\label{positive}
\end{equation}
Making  the same replacement in (\ref{B}), with the same error, we
obtain
\begin{eqnarray*}
& & -\frac{\beta\kappa^2}{2m}\int_0^{\infty}\frac{k^\prime}{m}
\exp\left\{-\frac{\beta k^{\prime 2}}{2m}-\zeta^ik^\prime\right\}dk\\
&=&  -\frac{\beta\kappa^2}{2m^2}M_1(\zeta^i).
\end{eqnarray*}
Therefore, the contribution coming from (\ref{B}) to the mass current is
\[-\frac{N_x\beta\kappa^2}{2m^2Z_i\epsilon}M_1(\zeta^i)=
-\frac{N_x\beta\kappa^2}{2m^2}\left(\frac{\beta}{2\pi m}\right)^{1/2}
e^{-m(\zeta^i)^2/2\beta}M_1(\zeta^i),
\]
which, to zeroth order in $\zeta^i$, gives
\begin{equation}
-N_x\left(\frac{\beta}{2\pi m}\right)^{1/2}\ell\partial_i\Phi(x)=
-\frac{N_x}{k_B\Theta^{1/2}}\frac{\ell c}{2\pi\Theta_0}\partial_i\Phi(x)
=-\frac{\lambda N_x}{\rho k_B\Theta^{1/2}}\partial_i\Phi(x).
\label{Ndrift}
\end{equation}
When multiplied by $m/a^3$ this is what we call the Smoluchowski, or drift,
current:
\begin{equation}
J^i_S=-{\lambda}\frac{\partial_i\Phi(x)}{k_B\Theta^{1/2}}.
\label{drift}
\end{equation}
The integral in (\ref{C}) is bounded by
\[-\frac{\kappa^2\zeta^i}{2m}M_0(\zeta^i).\]
Expanding $M_0$ to zeroth order in $\zeta^i$, this gives
\[-\frac{\zeta^i\kappa^2}{2m}\frac{1}{2}\left(\frac{2\pi m}{\beta}\right)^{1/2}
=-\frac{\zeta^i\ell\partial_i\Phi(x)}{2}\left(\frac{2\pi m}
{\beta}\right)^{1/2}=\left(\frac{\pi}{2}\right)^{1/2}\frac{\ell
u^i}{c}\partial_i\Phi ,\] so that it can be ignored in the limit.

We now turn our attention to the second term in the current (\ref{J}). We
can again replace the factor  $\ell r_-$
by $|k^i|/m$ (Appendix 2), so that the
contribution to the mass current from this
term is
\begin{equation}
\frac{N_x}{Z_i\epsilon}\int_{-\infty}^{0}\frac{k}{m}
\exp\left\{-\frac{\beta k^2}{2m}-\zeta^ik\right\}dk.
\label{negative}
\end{equation}
When we combine this with (\ref{positive}) and
multiply it all by $m/a^3$ what we find is simply $\rho u^i$.

Finally, we need to deal with the last term in (\ref{J}), which in
the limit becomes just $-\ell\partial_i F^i_x$. As we have just shown, the
only non-negligible terms in $F^i_x$ itself are (\ref{positive}) and
(\ref{Ndrift}). But (\ref{Ndrift}) is already of order $\ell c$ and therefore
can be ignored when multiplied by the additional $\ell$ above. The only
term that survives is
\begin{eqnarray*}
& & -\ell\partial_i\left[\frac{N_x}{Z_i\epsilon}\int_0^{\infty}\frac{k}{m}
\exp\left\{-\frac{\beta k^2}{2m}-\zeta^i k\right\}dk\right] \\
&=& -\ell\partial_i\left[\frac{N_x}{m}\left(\frac{\beta}{2\pi m}\right)^{1/2}
e^{-m(\zeta^i)^2/2\beta}M_1(\zeta^i)\right],
\end{eqnarray*}
which, to zeroth order in $\zeta^i$, is
\begin{equation}
-\ell\partial_i\left(\frac{N_x}{(2\pi m\beta)^{1/2}}\right)
=-\frac{\ell c}{(2\pi\Theta_0)^{1/2}}\partial_i\left(\frac{N_x}{k_B^{1/2}
\beta^{1/2}}\right)=-\frac{\lambda}{\rho}\partial_i(\Theta^{1/2}N_x).
\label{Nfick}
\end{equation}
When multiplied by $m/a^3$ this is what we call the diffusion current
\begin{equation}
J^i_d=-\frac{\lambda}{\rho}\partial_i(\Theta^{1/2}\rho),
\label{Fick}
\end{equation}
which is made up of the Fick current
\begin{equation}
-\lambda\Theta^{1/2}\partial_i(\log\rho)
\end{equation}
and the Soret current
\begin{equation}
-\lambda(2\Theta^{1/2})^{-1}\partial_i\Theta.
\end{equation}

Therefore, we obtain the total mass current by collecting together
(\ref{positive}), (\ref{negative}), (\ref{Ndrift}) and (\ref{Nfick}), that is
\begin{equation}
J^i_x=N_xu^i_x - \frac{\lambda}{\rho}\frac{N_x}{k_B\Theta^{1/2}}\partial_i
\Phi(x)-\frac{\lambda}{\rho}\partial_i(\Theta^{1/2}N_x).
\label{finalJ}
\end{equation}
We now go back to (\ref{iNchange}) and expand the finite difference in there
as
\[\frac{J^i_{x+\ell e_i}-J^i_x}{\ell}=\frac{\partial J^i_x}{\partial x^i}
+\frac{\ell}{2}\frac{\partial^2 J^i_x}{\partial x^{i^2}}+O(\ell^2).\]
Since the expression we just found for $J^i_x$ does not contain any term with
a large factor $c$, we see that, in the limit $\ell\rightarrow 0$
subject to keeping $\ell c$ finite, equation (\ref{Nchange}) becomes
\begin{equation}
\frac{\partial N_x}{\partial t}+\mbox{div}J=0.
\label{evolution1}
\end{equation}

Multiplying both sides of (\ref{evolution1}) by $m/a^3$ gives us
the equation for the time evolution of the particle's density
\begin{equation}
\frac{\partial\rho}{\partial t} + \mbox{div}({\bf u}\rho)
= \lambda\mbox{ div}\left(\rho^{-1}\nabla (\Theta^{1/2}\rho) +
\frac{\nabla\Phi}{k_B\Theta^{1/2}}\right),
\label{density1}
\end{equation}
or
\begin{equation}
\frac{\partial\rho}{\partial t} + \mbox{div}(J_{\rho})=0
\label{density2}
\end{equation}
where the conserved density current is found to be
\begin{equation}
J_{\rho}= {\bf u}\rho + J_d + J_S.
\label{density3}
\end{equation}

\subsection{Dynamics of the energy in an external potential}

Let $J_x^i/\ell$ be now the change in the value of $E_x$ due to the
hoppings occuring between $x$ and $x-\ell e_i$. As
before, the change due to exchanges with both $x\pm \ell e_i$
in an interval $\delta t$ is
\begin{equation}
\delta_i E_x = -\frac{J^i_{x+\ell e_i}-J^i_x}{\ell}\delta t
\label{iEchange}
\end{equation}
and the total change due in $E_x$ in an interval $\delta t$ due to
hoppings in all directions is
\begin{equation}
\delta E_x = (\delta_1 E_x + \delta_2 E_x + \delta_3 E_x)\delta t.
\label{Echange}
\end{equation}

We have that
\begin{equation}
J^i_x = -\sum_{k^i\leq 0} \ell r_-(k_x)p_x({\bf k}){\cal E}_x({\bf k}) +
\sum_{k^i\geq \kappa^i} \ell r_+(k_{x-\ell e_i})p_{x-\ell e_i}({\bf k})
{\cal E}_{x-\ell e_i}({\bf k}),
\label{homer}
\end{equation}
where ${\cal E}={\bf k}\cdot{\bf k}/2m+\Phi(x)$.

The analogue of the quantity $F_x$ of the previous section is now
\begin{equation}
G^i_x=N_x(Z\epsilon^3)^{-1}\int_{k^i\geq\kappa^i}\ell r_+(k^i)
\left(\frac{{\bf k\cdot k}}{2m}+\Phi(x)\right)
\exp\left\{-\frac{\beta{\bf k\cdot k}}{2m}
-\mbox{\boldmath$\zeta$}\cdot{\bf k}\right\}d^3{\bf k}.
\end{equation}

Adding and subtracting this to (\ref{homer}), replacing sums by integrals and
again using that $p_x(k)=N_x\bar{p}_x(k)$, we obtain
\begin{eqnarray}
J^i_x &=& G^i_x -\frac{N_x}{Z\epsilon^3}\int_{k^i\leq 0}\ell r_-(k_x)
\left(\frac{{\bf k\cdot k}}{2m}+\Phi(x)\right)
\exp\left\{-\frac{\beta{\bf k\cdot k}}{2m}
-\mbox{\boldmath$\zeta$}\cdot{\bf k}\right\}d^3{\bf k}\nonumber\\
&& -\ell\left(G^i_x-G^i_{x-\ell e_i}\right)/\ell.
\label{Jay}
\end{eqnarray}

We calculate $G_x$ first (for $i=1$). In the integral
\[\int_{k_1\geq\kappa}\frac{k_1+(k_1^2-\kappa^2)^{1/2}}{2m}
\left(\frac{{\bf k\cdot k}}{2m}+\Phi(x)\right)
\exp\left\{-\frac{\beta{\bf k\cdot k}}{2m}
-\mbox{\boldmath$\zeta$}\cdot{\bf k}\right\}d^3{\bf k},\]
we make the change of variables $k_1^{\prime 2}=k_1^2-\kappa^2$ while
keeping $k_2^{\prime}=k_2$ and $k_3^{\prime}=k_3$. Note that
\[k_1^2/2m+\Phi(x)=k_1^{\prime 2}/2m+\Phi(x+\ell e_1)\]
and $k_1dk_1=k_1^\prime dk_1^\prime$, so defining
\[A({\bf k^\prime})=\left(\frac{{\bf k^\prime\cdot k^\prime}}{2m}
+\Phi(x+\ell e_1)\right)
\exp\left\{-\frac{\beta{\bf k^\prime\cdot k^\prime}}{2m}
-\mbox{\boldmath$\zeta$}\cdot{\bf k^\prime}\right\},\]
the integral becomes
\begin{equation}
\int_{k_1^\prime\geq 0}
\frac{k_1^\prime \left[(k_1^{\prime 2}+\kappa^2)^{1/2}+k_1^\prime\right]}
{2m(k_1^{\prime 2}+\kappa^2)^{1/2}}A({\bf k^\prime})
\exp\left\{-\frac{\beta\kappa^2}{2m}-\left((k_1^{\prime 2}
+\kappa^2)^{1/2}-k_1^\prime\right)\zeta^1\right\}d^3{\bf k^\prime}
\end{equation}

We now expand the exponential to first
order
\[\exp\left\{-\frac{\beta\kappa^2}{2m}
-\left[\left(k_1^{\prime 2}+\kappa^2\right)^{1/2}-k_1^\prime\right]
\zeta^1\right\} = 1-\frac{\beta\kappa^2}{2m}
-\left[k_1^\prime-\left(k_1^{\prime 2}+\kappa^2\right)^{1/2}\right]\zeta^1.\]
Thus the integral splits in the following three terms
\begin{eqnarray}
& &\frac{N_x}{Z\epsilon^3}\int_{k_1^\prime\geq 0}
\frac{k_1^\prime \left[(k_1^{\prime 2}+\kappa^2)^{1/2}+k_1^\prime\right]}
{2m(k_1^{\prime 2}+\kappa^2)^{1/2}}
A({\bf k^\prime})d^3{\bf k^\prime}
\label{energyA}\\
&- &\frac{N_x\beta\kappa^2}{2mZ\epsilon^3}\int_{k_1^\prime\geq 0}
\frac{k_1^\prime \left[(k_1^{\prime 2}+\kappa^2)^{1/2}+k_1^\prime\right]}
{2m(k_1^{\prime 2}+\kappa^2)^{1/2}}
A({\bf k^\prime})d^3{\bf k^\prime}
\label{energyB}\\
& &-\frac{N_x\kappa^2}{Z\epsilon^3}\int_{k_1^\prime\geq 0}
\frac{k_1^\prime}
{2m(k_1^{\prime 2}+\kappa^2)^{1/2}}
A({\bf k^\prime})d^3{\bf k^\prime}.
\label{energyC}
\end{eqnarray}

If, we approximate the hopping rates
appearing above simply by
$k_1^\prime/m$ (Appendix 3) and use that
$\Phi(x+\ell e_1)=\Phi(x)+\ell\partial_1\Phi(x)$, the integral in
(\ref{energyA}) becomes
\begin{eqnarray}
& &\frac{N_x}{Z\epsilon^3}\int_{k_1^\prime\geq 0}
\frac{k_1^\prime}{m}
\left(\frac{{\bf k^\prime\cdot k^\prime}}{2m}+\Phi(x)\right)
\exp\left\{-\frac{\beta{\bf k^\prime\cdot k^\prime}}{2m}
-\mbox{\boldmath$\zeta$}\cdot{\bf k^\prime}\right\}d^3{\bf k^\prime}\nonumber\\
&+&\frac{N_x}{Z\epsilon^3}\int_{k_1^\prime\geq 0}
\frac{k_1^\prime}{m}\ell\partial_1\Phi(x)
\exp\left\{-\frac{\beta{\bf k^\prime\cdot k^\prime}}{2m}
-\mbox{\boldmath$\zeta$}\cdot{\bf k^\prime}\right\}d^3{\bf k^\prime}
\label{bart1}
\end{eqnarray}

The first part of (\ref{bart1}) is later
going to be combined with the integral over the negative values of $k_1$
appearing in the second term of (\ref{Jay}). As for the second part of
(\ref{bart1}), we have
\begin{eqnarray*}
& &\frac{N_x}{Z\epsilon^3}\int_{k_1^\prime\geq 0}
\frac{k_1^\prime}{m}\ell\partial_1\Phi(x)
\exp\left\{-\frac{\beta{\bf k^\prime\cdot k^\prime}}{2m}
-\mbox{\boldmath$\zeta$}\cdot{\bf k^\prime}\right\}d^3{\bf k^\prime}\\
&=& \frac{N_x\ell\partial_1\Phi(x)}{Z_1\epsilon m}\int_{k_1^\prime\geq 0}
k_1^\prime\exp\left\{-\frac{\beta k_1^{\prime 2}}{2m}-\zeta^1k_1\right\}
dk_1^\prime \\
&=& \frac{N_x\ell\partial_1\Phi(x)}{Z_1\epsilon m}M_1(\zeta^1)=
\frac{N_x\ell\partial_1\Phi(x)}{m}\left(\frac{\beta}{2\pi m}\right)^{1/2}
M_1(\zeta^1)e^{-\frac{m(\zeta^1)^2}{2\beta}}
\end{eqnarray*}
Expanding it to zeroth order in $\zeta^1$, we get
\begin{equation}
\frac{N_x\ell\partial_1\Phi(x)}{\beta}\left(\frac{\beta}{2\pi m}\right)^{1/2}
=\frac{\lambda}{\rho}N_x\Theta^{1/2}\partial_1\Phi(x).
\label{lisa}
\end{equation}

For the integral (\ref{energyB}), we again  approximate the hopping rate by
$k_1^\prime/m$ and use that
$\Phi(x+\ell e_1)=\Phi(x)+\ell\partial_1\Phi(x)$, to obtain
\begin{eqnarray}
& &-\frac{N_x\beta\kappa^2}{2m^2Z\epsilon^3}\int_{k_1^\prime\geq 0}
k_1^\prime\left(\frac{{\bf k^\prime\cdot k^\prime}}{2m}\right)
\exp\left\{-\frac{\beta{\bf k^\prime\cdot k^\prime}}{2m}
-\mbox{\boldmath$\zeta$}\cdot{\bf k^\prime}\right\}d^3{\bf k^\prime}\nonumber\\
&&-\frac{N_x\beta\kappa^2}{2m^2Z\epsilon^3}\int_{k_1^\prime\geq 0}
k_1^\prime(\Phi(x)+\ell\partial_1\Phi(x))
\exp\left\{-\frac{\beta{\bf k^\prime\cdot k^\prime}}{2m}
-\mbox{\boldmath$\zeta$}\cdot{\bf k^\prime}\right\}d^3{\bf k^\prime}
\label{bart2}
\end{eqnarray}
The first term above divides into three integrals. The first one is
\begin{eqnarray}
&&-\frac{\beta N_x\kappa^2}{2m^2Z\epsilon^3}\int_{k_1^\prime\geq 0}
\frac{k_1^{\prime 3}}{2m}
\exp\left\{-\frac{\beta{\bf k^\prime\cdot k^\prime}}{2m}
-\mbox{\boldmath$\zeta$}\cdot{\bf k^\prime}\right\}d^3{\bf k^\prime}\nonumber\\
&=& -\frac{\beta N_x\ell\partial_1\Phi}{mZ_1\epsilon}\frac{M_3(\zeta^1)}{2m}
\nonumber\\
&=&-N_x\left(\frac{\beta}{2\pi m}\right)^{1/2}\ell\partial_1\Phi(x) k_B\Theta;
\end{eqnarray}
the second one is
\begin{eqnarray}
&&-\frac{\beta N_x\kappa^2}{2m^2Z_1Z_2\epsilon^2}\int_{k_1^\prime\geq 0}
\frac{k_1^{\prime 3}}{2m}e^{-\frac{\beta k_1^2}{2m}-\zeta^1 k_1}dk_1
\int_{k_2}k_2^2 e^{-\frac{\beta k_2^2}{2m}-\zeta^2 k_2}dk_2
\nonumber\\
&=&-\frac{\beta N_x\ell\partial_1\Phi(x)}{mZ_1\epsilon}\frac{M_1(\zeta^1)}{2m}
\left[\frac{m^2(\zeta^2)^2}{\beta^2}+\frac{m}{\beta}\right] \nonumber\\
&=& -N_x\left(\frac{\beta}{2\pi m}\right)^{1/2}\ell\partial_1\Phi(x)
\left[\frac{m(u^2)^2}{2}+\frac{1}{2}k_B\Theta\right];
\end{eqnarray}
while the third one is
\begin{eqnarray}
&&-\frac{\beta N_x\kappa^2}{2m^2Z_1Z_3\epsilon^2}\int_{k_1^\prime\geq 0}
\frac{k_1^{\prime 3}}{2m}e^{-\frac{\beta k_1^2}{2m}-\zeta^1 k_1}dk_1
\int_{k_3}k_3^2 e^{-\frac{\beta k_3^2}{2m}-\zeta^3 k_3}dk_3
\nonumber\\
&=&-\frac{\beta N_x\ell\partial_1\Phi(x)}{mZ_1\epsilon}\frac{M_1(\zeta^1)}{2m}
\left[\frac{m^2(\zeta^3)^2}{\beta^2}+\frac{m}{\beta}\right] \nonumber\\
&=& -N_x\left(\frac{\beta}{2\pi m}\right)^{1/2}\ell\partial_1\Phi(x)
\left[\frac{m(u^3)^2}{2}+\frac{1}{2}k_B\Theta\right].
\end{eqnarray}
So the total contribution from the first term of (\ref{bart2}) is
\begin{equation}
-N_x\left(\frac{\beta}{2\pi m}\right)^{1/2}\ell\partial_1\Phi(x)
\left(\frac{m(u^2)^2}{2}+\frac{m(u^3)^2}{2}+2k_B\Theta\right).
\end{equation}
The terms involving the velocities disappear in the limit. The remaining
term is
\begin{equation}
-2\frac{\lambda}{\rho}N_x\Theta^{1/2}\partial_1\Phi(x).
\label{calvin}
\end{equation}

As for the second part of (\ref{bart2}) we have
\begin{equation}
-\frac{N_x\beta\kappa^2}{2m^2Z\epsilon^3}
\left[\Phi(x)+\ell\partial_1\Phi(x)\right]M_1(\zeta^1)=
-N_x\left(\frac{\beta}{2\pi m}\right)^{1/2}\ell\partial_1\Phi(x)
\left[\Phi(x)+\ell\partial_1\Phi(x)\right],
\end{equation}
of which only the first term survives in the limit, leaving us with
\begin{equation}
-\lambda\frac{\partial_1\Phi(x)}{k_B\Theta^{1/2}}\frac{N_x}{\rho}\Phi(x)
\label{hobbes}
\end{equation}

The integral in (\ref{energyC}) is itself of smaller order and can be
ignored (Appendix 4).
This completes the contribution of $G^i_x$ to the energy
current.

We move on to deal with the second term in (\ref{Jay}). The factor
$\ell r_-$ can be replaced by $-k_1/m$
(Appendix 5), leading to a contribution
of the form
\begin{equation}
\frac{N_x}{Z\epsilon^3}\int_{k_1\leq 0}
\frac{k_1}{m}
\left(\frac{{\bf k\cdot k}}{2m}+\Phi(x)\right)
\exp\left\{-\frac{\beta{\bf k\cdot k}}{2m}
-\mbox{\boldmath$\zeta$}\cdot{\bf k}\right\}d^3{\bf k}.
\end{equation}
As promised, this joins the first part of (\ref{bart1}) to give
\begin{eqnarray}
N_x{\bf E}_{\bar{p}}\left[\frac{{\cal P}_1}{m}{\cal E}_x\right] &=&
N_x{\bf E}_{\bar{p}}\left[\frac{{\cal P}_1}{m}\right]{\bf E}_{\bar{p}}
[{\cal E}]+N_x\overline{cor}
\left(\frac{{\cal P}^1}{m},{\cal E}_x\right) \nonumber\\
&=& u^1_xE_x + \frac{N_x}{m}\frac{\partial^2\log Z}
{\partial\zeta_1\partial\beta} \nonumber \\
&=& u^1_xE_x + N_xk_B\Theta_xu^1_x
\label{monica}
\end{eqnarray}

We finally look at the last part of (\ref{Jay}), which in the limit becomes
$-\ell\partial_1 G_x$. As we have just seen, all the contribution
coming from $G_x$ are already of order $\ell c$ (and can therefore
be discarded when multiplied by the additional $\ell$ above) with the
exception of the first part of (\ref{bart1}), that is,
\begin{equation}
\frac{N_x}{Z\epsilon^3}\int_{k_1^\prime\geq 0}
\frac{k_1^\prime}{m}
\left(\frac{{\bf k^\prime\cdot k^\prime}}{2m}+\Phi(x)\right)
\exp\left\{-\frac{\beta{\bf k^\prime\cdot k^\prime}}{2m}
-\mbox{\boldmath$\zeta$}\cdot{\bf k^\prime}\right\}d^3{\bf k^\prime}.
\end{equation}
We recognise the term not involving $\Phi$ as being equal to the first
term in (\ref{bart2}) times a factor
$-2m/\beta\kappa^2=-(\beta\ell\partial_1\Phi(x))^{-1}$. Therefore, its contribution to
the energy current is
\begin{equation}
-\ell\partial_1\left[2\frac{\lambda}{\ell\rho}N_xk_B\Theta^{3/2}\right]=
-2\frac{\lambda}{\rho}\partial_1(N_xk_B\Theta^{3/2}).
\label{cebolinha}
\end{equation}

As for the term involving $\Phi$, we again recognise it as being equal to
the term involving $\Phi$ in the second part of (\ref{bart2}) times
the factor $-2m/\beta\kappa^2=-1/\beta\ell\partial_1\Phi$. Therefore, its
contribution to the energy current is
\begin{equation}
-\ell\partial_1\left[\frac{\lambda}{\rho}\frac{N_x\Theta^{1/2}}{\ell}\Phi(x)
\right]
= -\frac{\lambda}{\rho}\partial_1(N_x\Theta^{1/2})\Phi(x)-
\frac{\lambda}{\rho}N_x\Theta^{1/2}\partial_1\Phi(x)
\label{cascao}
\end{equation}

We can now take a breath and collect all the terms we have obtained for
the energy current to put back in (\ref{Jay}). They are (\ref{lisa}),
(\ref{calvin}), (\ref{hobbes}), (\ref{monica}), (\ref{cebolinha}) and
(\ref{cascao}) and the end result is
\begin{eqnarray}
J^i_x &=& u^i_x(E_x + N_xk_B\Theta_x) - 2\frac{\lambda}{\rho}N_x\Theta^{1/2}
\partial_i\Phi(x) \nonumber \\
&-& \lambda\frac{\partial_i\Phi(x)}{k_B\Theta^{1/2}}\frac{N_x}{\rho}\Phi(x)
-2\frac{\lambda}{\rho}\partial_i(N_xk_B\Theta^{3/2})
-\frac{\lambda}{\rho}\partial_i(N_x\Theta^{1/2})\Phi(x).
\end{eqnarray}

Again we see that none of the terms in $J^i_x$ contains a large factor $c$,
so that the discussion preceding (\ref{evolution1}) applies here as well and
in the limit we obtain
\begin{equation}
\frac{\partial E_x}{\partial t}+\mbox{div}J=0,
\label{evolution2}
\end{equation}
that is
\begin{eqnarray}
&&\frac{\partial E_x}{\partial t} + \mbox{ div }[{\bf u}_x(E_x +
N_xk_B\Theta_x)] = \mbox{ div }\left[2\frac{\lambda}{\rho}N_x\Theta^{1/2}
\nabla\Phi(x) \right.\nonumber \\
&+&\left.\lambda\frac{\nabla\Phi(x)}{k_B\Theta^{1/2}}\frac{N_x}{\rho}\Phi(x)
+2\frac{\lambda}{\rho}\nabla(N_xk_B\Theta^{3/2})
+\frac{\lambda}{\rho}\nabla(N_x\Theta^{1/2})\Phi(x)\right].
\end{eqnarray}

We now use that $e(x)=E_x/(mN_x)$, $\phi(x)=\Phi(x)/m$, $\rho(x)=mN_x/a^3$
and $P_x=N_xK_B\Theta_x/a^3$, divide both sides of the previous equation by
$a^3$ and obtain
\begin{eqnarray}
\frac{\partial(\rho e)}{\partial t} +  \mbox{ div }[{\bf u}(\rho e +
P)] &=& \lambda\mbox{ div }\left[2\rho^{-1}\nabla\left(\Theta^{1/2}P\right)
+ \rho^{-1}\nabla\left(\Theta^{1/2}\rho\right)\phi \right.\nonumber \\
&&+ \left.\frac{\nabla\Phi}{k_B\Theta^{1/2}}\phi + 2\frac{\nabla\Phi}
{k_B\Theta^{1/2}}P \right],
\label{energy1}
\end{eqnarray}
which can be written as (recall (\ref{drift}) and (\ref{Fick}))
\begin{equation}
\frac{\partial(\rho e)}{\partial t} +  \mbox{ div }[{\bf u}(\rho e +
P)+(J_d+J_S)\phi + 2J_S P] = 2\lambda\mbox{ div }\left[\rho^{-1}\nabla
\left(P\Theta^{1/2}\right)\right]
\label{energy2}
\end{equation}

\subsection{Dynamics of the momentum in an external field}

Since momentum is not conserved (as there are body-forces due to
the external field), the rate of change of momentum density will
not be the divergence of something; we expect the extra term to be
$\rho\,{\bf f}$ where ${\bf f}:=-\nabla\Phi/m$ is the force per
unit mass. To see this, let us define the current $J_x^i$ as in
the previous sections, namely
\begin{equation}
J^i_x = -\sum_{k^i\leq 0} \ell r_-(k^i_x)p_x(k){\cal P}^j_x(k) +
\sum_{k^i\geq \kappa^i} \ell r_+(k^i_{x-\ell e_i})p_{x-\ell e_i}(k)
{\cal P}^j_{x-\ell e_i}(k)
\label{mhomer}
\end{equation}

Then the change in $\varpi^j_x$ due to exchanges with both $x\pm\ell e_i$
in an interval $\delta t$ will only be given by the usual
\begin{equation}
\delta_i \varpi^j_x = -\frac{J^i_{x+\ell e^i}-J^i_x}{\ell}\delta t
\label{iMchange}
\end{equation}
for $i\neq j$, because it is implicit in this formula that the particles
hopping from  $x \pm \ell e_i$ to $x$ have their $j$-component of the momentum
unchanged during the jump. We do this case first. The analogue of
$F^i_x$ and $G^i_x$ from the previous
sections is now
\begin{equation}
H^i_x=N_x(Z\epsilon^3)^{-1}\int_{k^i\geq\kappa^i}\ell r_+(k^i)k^j
\exp\left\{-\frac{\beta{\bf k\cdot k}}{2m}
-\mbox{\boldmath$\zeta$}\cdot{\bf k}\right\}d^3{\bf k}.
\label{dennis}
\end{equation}

Adding and subtracting this to (\ref{mhomer}), replacing sums by integral and
using that $p_x(k)=N_x\bar{p}_x(k)$, we obtain the familiar form
\begin{eqnarray}
J^i_x&=&H^i_x -\frac{N_x}{Z\epsilon^3}\int_{k^i\leq 0}\ell r_-(k^i)k^j
\exp\left\{-\frac{\beta{\bf k\cdot k}}{2m}
-\mbox{\boldmath$\zeta$}\cdot{\bf k}\right\}d^3{\bf k}\nonumber\\
&-&\ell\left(H^i_x-H^i_{x-\ell e_i}\right)/\ell
\label{jay}.
\end{eqnarray}

Notice that since $\varpi^j_x=mN_xu^j_x$, in this section we shall keep
all terms of order $mu^j_x$. If we now perform in (\ref{dennis})
the integrations over $k_r$ ($r\neq i$,
$r\neq j$) and  $k_j$ we find
\begin{eqnarray*}
H^i_x &=& mu^j_x\frac{N_x}{Z_i\epsilon}\int_{k\geq\kappa}\ell r_+(k)
\exp\left\{-\frac{\beta k^2}{2m}-\zeta k\right\}dk \\
&=& mu^j_xF^i_x.
\end{eqnarray*}
But now we can use the calculation we have already done for $F^i_x$, that is,
in the limit we have
\begin{eqnarray}
H^i_x &=& mu^j_x\frac{N_x}{Z_i\epsilon}\int_0^{\infty}\frac{k^\prime}{m}
\exp\left\{-\frac{\beta k^{\prime 2}}{2m}-\zeta^ik^\prime\right\}dk
\nonumber\\
&-& \left[\frac{\lambda N_x}{\rho k_B\Theta^{1/2}}\partial_i\Phi(x)\right]
mu^j_x
\label{dennis2}
\end{eqnarray}
The second term above, when multiplied by $m/a^3$, simply gives
$mu^j_xJ^i_S$.

As for the second term in (\ref{jay}), we can also perform the integrations
over $k_r$ ($r\neq i$,$r\neq j$) and  $k_j$, as well as to replace
the factor $\ell r_-(k^i)$ by $-k^i/m$ to find
\begin{equation}
mu^j_x\frac{N_x}{Z_i\epsilon}\int_{-\infty}^{0}\frac{k}{m}
\exp\left\{-\frac{\beta k^2}{2m}-\zeta^ik\right\}dk.
\end{equation}
When we add this to what we have found in the first term in (\ref{dennis2})
and multiply them $m/a^3$, the resulting term is $m\rho u^j_x u^i_x$.

Finally, we look at the last term in (\ref{jay}), which in the limit becomes
$-\ell\partial_iH^i_x$. Since the second term in (\ref{dennis2}) is already
of order $\ell c$, we see that the only surviving contribution here is
\begin{equation}
-\ell\partial_i\left[mu^j_x\frac{N_x}{Z_i\epsilon}\int_0^{\infty}
\frac{k^\prime}{m}
\exp\left\{-\frac{\beta k^{\prime 2}}{2m}-\zeta^ik^\prime\right\}dk\right]
=-\ell\partial_i\left[mu^j_x\frac{N_x}{Z_i\epsilon m}M_1(\zeta^i)\right],
\end{equation}
which, to zeroth order in $\zeta^i$, gives
\begin{eqnarray}
-\ell\partial_i\left[mu^j_x\frac{N_x}{m}\left(\frac{\beta}{2\pi m}\right)^{1/2}
\frac{m}{\beta}\right]&=&-\frac{\ell c}{(2\pi \Theta_0)^{1/2}}\partial_i
(N_x\Theta^{1/2}mu^j)\nonumber\\
&=&-\frac{\lambda}{\rho}\partial_i(N_x\Theta^{1/2} mu^j).
\label{pedro}
\end{eqnarray}

Therefore, the momentum current, for $i\neq j$, is given by
\begin{equation}
J^i_x=mN_x u^j_x u^i_x-\left[\frac{\lambda N_x}{\rho k_B\Theta^{1/2}}
\partial_i\Phi(x)\right]mu^j_x -\frac{\lambda}{\rho}
\partial_i(N_x\Theta^{1/2}mu^j).
\label{icurrent}
\end{equation}
Once more, none of the above terms contain a large factor $c$, so in the
limit $\ell\rightarrow 0$ subject to $\ell c$ finite, we can approximate
the term $(J^i_{x+\ell e_i}-J^i_x)/\ell$ in (\ref{iMchange})
simply by $\partial_i J^i_x$.

For $i=j$, equation (\ref{iMchange}) is {\em not} correct. In this case,
if $k_{x-\ell e_j}>\kappa^j_{x-\ell e_j}$
is the $j$-component of the
momentum for a particle at $x-\ell e_j$, then
\[k^\prime_{x-\ell e_j}=[k_{x-\ell e_j}^2-2m\ell\partial_j
\Phi(x-\ell e_j)]^{1/2}\]
is the $j$-component of its momentum when it arrives at $x$. Similarly,
if $k_{x+\ell e_j}<0$ is the $j$-component of the momentum for a
particle at $x+\ell e_j$, then
\[k^{\prime\prime}_{x+\ell e_j}=-[k_{x+\ell e_i}^2+2m\ell\partial_j
\Phi(x)]^{1/2}\]
is the $j$-component of its momentum
when it gets to $x$. Therefore, the total rate of change in
$\varpi^j_x$ due to exchanges with
$x\pm\ell e_j$  consists of the following four terms
\begin{eqnarray}
\frac{\delta_j \varpi^j_x}{\delta t} &=&\sum_{k_{x+\ell e_j}\leq 0}
r_-(k_{x+\ell e_j})
k^{\prime\prime}_{x+\ell e_j} p_{x+\ell e_j}({\bf k}_{x+\ell e_j})\\
&-&\sum_{k_x\geq \kappa^j_x}r_+(k_x)k_x p_x({\bf k}_x)
-\sum_{k_x\leq 0}r_-(k_x)k_x p_x({\bf k}_x)\\
&+&\sum_{k_{x-\ell e_j}\geq\kappa^j_{x-\ell e_j}} r_+(k_{x-\ell e_j})
k^\prime_{x-\ell e_j} p_{x-\ell e_j}({\bf k}_{x-\ell e_j})
\end{eqnarray}

If we now recall from (\ref{rates}) what the hopping rates look like,
and use that
\begin{eqnarray*}
\left(k_{x+\ell e_j}+k_{x+\ell e_j}^{\prime\prime}\right)
k_{x+\ell e_j}^{\prime\prime}
&=&\left(k_{x+\ell e_j}+k_{x+\ell e_j}^{\prime\prime}\right)
k_{x+\ell e_j}+2m\ell\partial_j\Phi(x)\\
\left(k_{x-\ell e_j}+k_{x-\ell e_j}^\prime\right)k_{x-\ell e_j}^\prime
&=&\left(k_{x-\ell e_j}+k_{x-\ell e_j}^\prime\right)k_{x-\ell e_j}-2\ell
m\partial_j\Phi(x-\ell e_j),
\end{eqnarray*}
then we can rewrite the above as
\begin{eqnarray*}
\frac{\delta_j \varpi^j_x}{\delta t} &=&\sum_{k_{x+\ell e_j}\leq 0}
r_-(k_{x+\ell e_j})
k_{x+\ell e_j} p_{x+\ell e_j}({\bf k}_{x+\ell e_j})\\
&-&\sum_{k_x\geq \kappa^j_x}r_+(k_x)k_x p_x({\bf k}_x)
-\sum_{k_x\leq 0}r_-(k_x)k_x p_x({\bf k}_x)\\
&+&\sum_{k_{x-\ell e_j}\geq\kappa^j_{x-\ell e_j}}r_+(k_{x-\ell e_j})
k_{x-\ell e_j} p_{x-\ell e_j}({\bf k}_{x-\ell e_j})\\
&-& \partial_j\Phi(x)\sum_{k_{x+\ell e_j}\leq 0}
p_{x+\ell e_j}({\bf k}_{x+\ell e_j})\\
&-&\partial_j\Phi(x-\ell e_j)
\sum_{k_{x-\ell e_j}\geq\kappa^j_{x-\ell e_j}}
p_{x-\ell e_j}({\bf k}_{x-\ell e_j})
\end{eqnarray*}
which we then recognise as
\begin{eqnarray}
\frac{\delta_j \varpi^j_x}{\delta t} &=& -\frac{J^j_{x+\ell e^j}-J^j_x}
{\ell} \nonumber\\
&-& \partial_j\Phi(x)\sum_{k_{x+\ell e_j}\leq 0}
p_{x+\ell e_j}({\bf k}_{x+\ell e_j}) \nonumber\\
&-&\partial_j\Phi(x-\ell e_j)
\sum_{k_{x-\ell e_j}\geq\kappa^j_{x-\ell e_j}}
p_{x-\ell e_j}({\bf k}_{x-\ell e_j}).
\label{jMchange}
\end{eqnarray}
In the limit, the last two terms above add up to
$N_x\partial_j\Phi(x)$ (Appendix 6).

Let us now calculate $J^j_x$, that is
\begin{eqnarray}
J^j_x&=&H^j_x -\frac{N_x}{Z\epsilon^3}\int_{k^j\leq 0}\ell r_-(k^j)k^j
\exp\left\{-\frac{\beta{\bf k\cdot k}}{2m}
-\mbox{\boldmath$\zeta$}\cdot{\bf k}\right\}d^3{\bf k}\nonumber\\
&-&\ell\left(H^j_x-H^j_{x-\ell e_j}\right)/\ell
\label{jjay}.
\end{eqnarray}
We find that $H^j_x$ reduces to
\[H^j_x=\frac{N_x}{Z_j\epsilon}\int_{k\geq\kappa}
\frac{k+(k^2-\kappa^2)^{1/2}}{2m}k
\exp\left\{-\frac{\beta k^2}{2m}-\zeta^jk\right\}dk,\]
where we can make the familiar change of variables
$k^{\prime 2}=k^2-\kappa^2$ to obtain
\begin{eqnarray*}
\lefteqn{H^j_x=\frac{N_x}{Z_j\epsilon}\int_{k^\prime\geq 0}
\left[\frac{(k^{\prime 2}+\kappa^2)^{1/2}+k^\prime}{2m}\right]
k^\prime
\exp\left\{-\frac{\beta k^{\prime 2}}{2m}-\zeta^jk^\prime\right\}}\\
&\times &\exp\left\{-\frac{\beta\kappa^2}{2m}-\left((k^{\prime 2}
+\kappa^2)^{1/2}-k^\prime\right)\zeta^j\right\}dk^\prime.
\end{eqnarray*}
Expanding the exponential to first order gives the usual three terms
\begin{eqnarray}
H^j_x &=& \frac{N_x}{Z_j\epsilon}\left(\int_{k^\prime\geq 0}
\left[\frac{(k^{\prime 2}+\kappa^2)^{1/2}+k^\prime}{2m}\right]k^\prime
\exp\left\{-\frac{\beta k^{\prime 2}}{2m}-\zeta^j k^\prime\right\}dk^\prime
\right.
\label{mA}\\
& &-\frac{\beta\kappa^2}{2m}\int_{k^\prime\geq 0}
\left[\frac{(k^{\prime 2}+\kappa^2)^{1/2}+k^\prime}{2m}\right]k^\prime
\exp\left\{-\frac{\beta k^{\prime 2}}{2m}-\zeta^jk^\prime\right\}dk^\prime
\label{mB}\\
& &\left.-\frac{\kappa^2\zeta^j}{2m}\int_{k^\prime\geq 0}k^\prime
\exp\left\{-\frac{\beta k^{\prime 2}}{2m}-\zeta^jk^\prime\right\}dk^\prime.
\right)
\label{mC}
\end{eqnarray}

In (\ref{mA}), we can replace the factor
\[\frac{(k^{\prime 2}+\kappa^2)^{1/2}+k^\prime}{2m}\]
by $k^\prime/m$ (Appendix 7), so that it amounts to
\begin{equation}
\frac{N_x}{Z_j\epsilon m}\int_{k^\prime\geq 0}(k^\prime)^2
\exp\left\{-\frac{\beta k^{\prime 2}}{2m}-\zeta^j k^\prime\right\}dk^\prime.
\label{ana}
\end{equation}

In (\ref{mB}), the same replacement gives
\begin{equation}
-\frac{N_x\beta\kappa^2}{2Z_j\epsilon m^2}\int_{k^\prime\geq 0}(k^\prime)^2
\exp\left\{-\frac{\beta k^{\prime 2}}{2m}-\zeta^j k^\prime\right\}dk^\prime =
-\frac{N_x\beta\kappa^2}{2Z_j\epsilon m^2}M_2(\zeta^j),
\end{equation}
which, to first order in $\zeta_j$ (recall that we are keeping terms
proportional to $mu^j$ in this section, since $\varpi^j_x=N_xmu^j_x$), gives
\begin{eqnarray}
&-&\frac{2N_x\beta m\ell\partial_j\Phi(x)}{2m^2}\left(\frac{\beta}{2\pi m}
\right)^{1/2}
\left[\left(\frac{\pi}{2}\right)^{1/2}\left(\frac{m}{\beta}\right)^{3/2}
-2\left(\frac{m}{\beta}\right)^2\zeta^j\right] = \nonumber\\
&-&\frac{\ell N_x \partial_j\Phi(x)}{2}-2N_x\left(\frac{\beta}{2\pi m}
\right)^{1/2}\ell\partial_j\Phi(x) mu^j
\end{eqnarray}
of which only the second term survives in the limit, resulting in
\begin{equation}
-2\left[\frac{\lambda N_x}{\rho k_B\Theta^{1/2}}\partial_j\Phi(x)\right]mu^j_x.
\label{fred}
\end{equation}

Similarly in (\ref{mC}), we  obtain
\begin{eqnarray}
-\frac{N_x\kappa^2\zeta^j}{2Z_j\epsilon m}M_1(\zeta^j) &=&
\frac{2N_x m \ell\partial_j\Phi\beta u^j_x}{2m}\left(\frac{\beta}{2\pi m}
\right)^{1/2}\frac{m}{\beta} \nonumber\\
&=& N_x\left(\frac{\beta}{2\pi m}\right)^{1/2}\ell\partial_j\Phi(x)
mu^j_x \nonumber\\
&=&\left[\frac{\lambda N_x}{\rho k_B\Theta^{1/2}}\partial_j\Phi(x)
\right]mu^j_x.
\label{wilma}
\end{eqnarray}

Moving to the second term in (\ref{jjay}), we replace the factor
$\ell r_-(k^j)$
by $-k^j/m$ (Appendix 8), so it contributes with
\begin{equation}
\frac{N_x}{Z_j\epsilon m}\int_{k\leq 0} k^2
\exp\left\{-\frac{\beta k^2}{2m}-\zeta^j k\right\}dk.
\label{matheus}
\end{equation}
Combining (\ref{ana}) and (\ref{matheus}), what we obtain is
\begin{eqnarray}
\frac{N_x}{mZ_j\epsilon}\int_{-\infty}^{\infty}(k^j)^2
\exp\left\{-\frac{\beta k^2}{2m}-\zeta^jk\right\}dk^j &=&\frac{N_x}{m}
E_{\bar{p}}\left[(k^j)^2\right] \nonumber \\
&=& \frac{N_x}{m}E_{\bar{p}}\left[k^j\right]^2 + \frac{N_x}{m}
\frac{\partial^2 \log Z_j}{\partial (\zeta^j)^2} \nonumber\\
&=& mN_x(u^j_x)^2 + N_xk_B\Theta_x.
\label{leonardo}
\end{eqnarray}

Finally, for the last term in (\ref{jjay}), which in the limit becomes
$-\ell\partial_j H^j_x$, only (\ref{ana}) contributes, since the other terms in
$H^j_x$ are already of order $\ell c$. We get
\begin{eqnarray*}
-\ell\partial_jH^j_x &=& -\ell\partial_j\left[\frac{N_x}{Z_j\epsilon m}
\int_{k^\prime\geq 0}(k^\prime)^2
\exp\left\{-\frac{\beta k^{\prime 2}}{2m}-\zeta^j k^\prime\right\}
dk^\prime \right] \nonumber \\
&=&-\ell\partial_j\left[\frac{N_x}{Z_j\epsilon m}M_2(\zeta^j)\right],
\end{eqnarray*}
which, to first order in $\zeta^j$, gives
\begin{eqnarray}
&-&\ell\partial_j\left\{\frac{N_x}{m}\left(\frac{\beta}{2\pi m}
\right)^{1/2}
\left[\left(\frac{\pi}{2}\right)^{1/2}\left(\frac{m}{\beta}\right)^{3/2}
-2\left(\frac{m}{\beta}\right)^2\zeta^j\right]\right\} = \nonumber \\
&& -\frac{\ell}{2}\partial_j\left(N_xk_B\Theta_x\right)
- 2\ell\partial_j\left[mu^j N_x\left(\frac{\beta}{2\pi m}\right)^{1/2}
\frac{1}{\beta}\right] = \nonumber \\
&&-\frac{\ell}{2}\partial_j\left(N_xk_B\Theta_x\right)
-2\frac{\lambda}{\rho}\partial_j\left(N_x\Theta^{1/2}mu^j_x\right)
\label{barney}
\end{eqnarray}

Therefore, the momentum current, for $i=j$ is given by the sum of
(\ref{fred}), (\ref{wilma}), (\ref{leonardo}) and (\ref{barney}), that is
\begin{eqnarray*}
J^j_x = mN_x(u^j_x)^2 &+& N_xk_B\Theta_x-\left[\frac{\lambda N_x}
{\rho k_B\Theta^{1/2}}\partial_j\Phi(x)\right]mu^j_x\\
&-&\frac{\ell}{2}\partial_j\left(N_xk_B\Theta_x\right)
-2\frac{\lambda}{\rho}\partial_j\left(N_x\Theta^{1/2}mu^j_x\right).
\end{eqnarray*}

We see that the term $N_xk_B\Theta_x$ above is of a larger order than
the others and is not negligible in the expansion of the finite difference
$\left(J_{x+\ell e_j}-J_x\right)/\ell$. What we obtain is
\begin{eqnarray*}
\frac{J_{x+\ell e_j}-J_x}{\ell}&=&\frac{\partial J^j_x}{\partial x^j}
+ \frac{\ell}{2}\frac{\partial^2 J^j_x}{\partial x^{j^2}} + O(\ell^2) \\
&=& \frac{\partial J^j_x}{\partial x^j}+
\frac{\ell}{2}\frac{\partial^2 (N_xk_B\Theta_x)}{\partial x^{j^2}} + O(\ell^2)
\end{eqnarray*}
Thus, in this case, the finite difference
$\left(J_{x+\ell e_j}-J_x\right)/\ell$ can be
approximated by
\begin{equation}
\partial_j\left\{mN_x(u^j_x)^2 + N_xk_B\Theta_x-\left[\frac{\lambda N_x}
{\rho k_B\Theta^{1/2}}\partial_j\Phi(x)\right]mu^j_x
-2\frac{\lambda}{\rho}\partial_j\left(N_x\Theta^{1/2}mu^j_x\right)\right\}.
\end{equation}

Therefore, collecting together the contributions from $i=1,2,3$, we see
from (\ref{iMchange}), (\ref{jMchange}) and the equation above, that the
change in $\varpi^j$ is governed by the equation
\begin{eqnarray}
&&\frac{\partial\varpi^j_x}{\partial t}=\frac{\partial}{\partial x^j}\left[
\frac{\lambda}{\rho}\frac{\partial}{\partial x^j}
\left(N_x\Theta^{1/2}mu^j_x\right)\right]-N_x\partial_j\Phi(x)
-\sum_{i=1}^3\frac{\partial}{\partial x^i}\left[mN_xu^i_xu^j_x\right.
\nonumber \\
&+&\left.N_xk_B\Theta_x\delta_{ij}-\left(\frac{\lambda N_x}{\rho k_B
\Theta^{1/2}}\partial_i\Phi\right)mu^j_x-\frac{\lambda}
{\rho}\frac{\partial}{\partial x^i}\left(N_x\Theta^{1/2}mu^j_x\right)\right]
\label{maux}
\end{eqnarray}

The first term above is not covariant and we need to average it over $SO(3)$.
The averaging procedure is explained in \cite{streater4}, and its result is
\begin{equation}
\sum_{i=1}^3\frac{\partial}{\partial x^i}\frac{\lambda}{5\rho}
\left[\frac{\partial}{\partial x^i}\left(N_x\Theta^{1/2}mu^j_x\right)+
2\frac{\partial}{\partial x^j}\left(N_x\Theta^{1/2}mu^i_x\right)\right].
\end{equation}

We can now put it back into (\ref{maux}), divide both sides of it by $a^3$
and use that $\varpi^j_x=mN_xu^j_x$, $\rho(x)=mN_x/a^3$,
$f^j_x=-\partial_j\Phi(x)/m$ and $P_x=N_xk_B\Theta_x/a^3$ to obtain
\begin{eqnarray}
\frac{\partial\rho(x)u^j_x}{\partial t}&=&\rho(x)f^j_x
-\sum_{i=1}^3\frac{\partial}{\partial x^i}\left\{\rho(x)u^i_xu^j_x
+ P_x\delta_{ij}-\frac{\lambda\partial_i\Phi}{k_B\Theta^{1/2}}u^j_x
\right.\nonumber \\
&-&\left.\frac{2\lambda}
{5\rho}\left[3\frac{\partial}{\partial x^i}\left(\rho(x)\Theta^{1/2}u^j_x
\right)+\frac{\partial}{\partial x^j}\left(\rho(x)\Theta^{1/2}u^i_x
\right)\right]\right\}.
\label{maux2}
\end{eqnarray}

In vector notation, this reads
\begin{eqnarray}
\frac{\partial\rho{\bf u}}{\partial t}&+&\mbox{div}(\rho{\bf u}\otimes{\bf u})
=\rho{\bf f}-\nabla P +\lambda\mbox{ div}\left(\frac{\nabla\Phi}
{k_B\Theta^{1/2}}\otimes{\bf u}\right) \nonumber \\
&+&\frac{2\lambda}{5}\partial_i\rho^{-1}\left[3\partial^i\left(\rho(x)
\Theta^{1/2}{\bf u}\right)
+\nabla\left(\rho(x)\Theta^{1/2}u^i\right)\right],
\label{mom1}
\end{eqnarray}
which can be written as
\begin{eqnarray}
\frac{\partial\rho{\bf u}}{\partial t}&+&\mbox{div}\left(\rho{\bf u}\otimes
{\bf u}+J_S\otimes{\bf u}\right)=\rho{\bf f}-\nabla P  \nonumber \\
&+&\frac{2\lambda}{5}\partial_i\rho^{-1}\left[3\partial^i\left(\rho(x)
\Theta^{1/2}{\bf u}\right)
+\nabla\left(\rho(x)\Theta^{1/2}u^i\right)\right].
\label{mom2}
\end{eqnarray}

\section{Appendices}

\subsection{Appendix 1}

The difference between the cases with and without an external potential
is
\begin{eqnarray*}
\frac{k^\prime}{m}-\frac{k^\prime
\left[(k^{\prime 2}+\kappa^2)^{1/2}+k^\prime\right]}
{2m(k^{\prime 2}+\kappa^2)^{1/2}} &=&
\frac{k^\prime \left[(k^{\prime 2}+\kappa^2)^{1/2}-k^\prime\right]}
{2m(k^{\prime 2}+\kappa^2)^{1/2}}\\
&\leq& \frac{(k^{\prime 2}+\kappa^2)^{1/2}-k^\prime}{2m}.
\end{eqnarray*}

So we need to bound integrals of the form
\[
B_1:=\frac{N_x}{Z_i\epsilon}\int_{k^\prime\geq 0}
\frac{(k^{\prime 2}+\kappa^2)^{1/2}-k^\prime}{2m}
\exp\left\{-\frac{\beta k^{\prime 2}}{2m}-\zeta^i k^\prime \right\}dk^\prime.\]
We have
\begin{eqnarray*}
(k^{\prime 2}+\kappa^2)^{1/2}-k^\prime &=&
\frac{\kappa^2}{k^\prime+(k^{\prime 2}+\kappa^2)^{1/2}}\\
&\leq& \left\{\begin{array}{cl}
\frac{\kappa^2}{2k^\prime}&\mbox{if }k^\prime\geq\kappa\\
\kappa&\mbox{if }0\leq k^\prime\leq\kappa.
\end{array}
\right.
\end{eqnarray*}

Hence
\begin{eqnarray*}
B_1 &\leq& \frac{N_x}{2m Z_i\epsilon}
\left(\int_0^\kappa \kappa \exp\left\{
-\frac{\beta k^{\prime 2}}{2m}-\zeta^i k^\prime \right\} dk^\prime \right.\\
&& \left. +\kappa^2\int_\kappa^\infty\frac{dk^\prime}{k^\prime}
\exp\left\{-\frac{\beta k^{\prime 2}}{2m}-\zeta_i k^\prime\right\}\right)\\
&\leq& \frac{N_x}{2m Z_i\epsilon}\left(\kappa^2
+\kappa^2\int_\kappa^\infty \frac{dk^\prime}{k^\prime}
\exp\left\{-\frac{\beta k^{\prime 2}}{2m}\right\}\right)\\
&\leq& \frac{N_x}{2m Z_i\epsilon}
\left(\kappa^2 - \kappa^2\log\left[\left(
\frac{\beta}{m}\right)^{1/2}\kappa\right]+\kappa^2\int_1^\infty \frac{dy}{y}
\exp\{1/2 y^2\}dy\right).
\end{eqnarray*}

Since $\kappa^2=2\ell m\partial\Phi$ and the term $N_x/(Z_i\epsilon)$ is
$O(1)$, we see that the error is $O(\ell\log\ell)$.

\subsection{Appendix 2}

Here the difference between the cases with and without an external potential is
\begin{equation*}
\frac{k}{m}-\frac{k-(k^2+\kappa^2)^{1/2}}{2m} =
\frac{k+(k^2+\kappa^2)^{1/2}}{2m},
\end{equation*}
so the integral we need to bound is
\[B_2:=\frac{N_x}{Z_i\epsilon}\int_{k\leq 0}\frac{(k^2+\kappa^2)^{1/2}+k}{2m}
\exp\left\{-\frac{\beta k^2}{2m}-\zeta^i k \right\}dk.\]

With the change of variable $k^\prime=-k$, we are led to the problem
of finding an upper bound for
\[\frac{N_x}{Z_i\epsilon}\int_{k^\prime\geq 0}
\frac{(k^{\prime 2}+\kappa^2)^{1/2}-k^\prime}{2m}
\exp\left\{-\frac{\beta k^{\prime 2}}{2m}+\zeta^i k^\prime \right\}dk^\prime.\]

But this reduces to the case of Appendix 1, since the quadratic
(negative) term in the exponent eventually (and in fact very quickly, since
$m$ is so small) overcomes the linear (positive) one.

\subsection{Appendix 3}

As in Appendix 1, the difference between the cases with or without an
external potential is
\begin{eqnarray*}
\frac{k_1^\prime}{m}-\frac{k_1^\prime
\left[(k_1^{\prime 2}+\kappa^2)^{1/2}+k_1^\prime\right]}
{2m(k_1^{\prime 2}+\kappa^2)^{1/2}} &=&
\frac{k_1^\prime \left[(k_1^{\prime 2}+\kappa^2)^{1/2}-k_1^\prime\right]}
{2m(k_1^{\prime 2}+\kappa^2)^{1/2}}\\
&\leq& \frac{(k_1^{\prime 2}+\kappa^2)^{1/2}-k_1^\prime}{2m}.
\end{eqnarray*}
so that we need to bound the integral
\begin{eqnarray*}
B_3 &:=& \frac{N_x}{Z\epsilon^3}\int_{k_1^\prime\geq 0}
\frac{(k_1^{\prime 2}+\kappa^2)^{1/2}-k_1^\prime}{2m}
A({\bf k^\prime})d^3{\bf k^\prime}. \nonumber \\
&=& \frac{N_x}{Z\epsilon^3}\int_{k_1^\prime\geq 0}d{\bf k^\prime}
\frac{(k_1^{\prime 2}+
\kappa^2)^{1/2}-k_1^\prime}
{2m} \\
&& \times \left(\frac{{\bf k^\prime\cdot k^\prime} }{2m}
+\Phi(x^\prime)\right)
\exp\left\{-\frac{\beta{\bf k^\prime\cdot k^\prime}}{2m}
-\mbox{\boldmath$\zeta$}\cdot{\bf k^\prime}\right\}
\end{eqnarray*}
which gives rise to the following four terms
\begin{eqnarray*}
& & \frac{N_x}{Z_1\epsilon}\int_{k_1^\prime\geq 0}
\left[\frac{(k_1^{\prime 2}+\kappa^2)^{1/2}-k_1^\prime}{2m}\right]
\frac{k_1^{\prime 2}}{2m}\exp\left\{-\frac{\beta k_1^{\prime 2}}{2m}-\zeta^i
k_1^\prime \right\}dk_1^\prime \\
&+& \frac{N_x}{(Z_1\epsilon)(2mZ_2\epsilon)}B_1[M_2(\zeta^2)+M_2(-\zeta^2)] \\
&+& \frac{N_x}{(Z_1\epsilon)(2mZ_3\epsilon)}B_1[M_2(\zeta^3)+M_2(-\zeta^3)] \\
&+& \frac{N_x}{Z_1\epsilon}\Phi(x^\prime)B_1
\end{eqnarray*}

Using that, to zeroth order in $\zeta$,  $M_2$ and $Z_i\epsilon$
are proportional, respectively, to
$m^{3/2}$ and $m^{-1/2}$,
we conclude that the last three terms are all of the same order
as $B_1$, that is, $O(\ell\log\ell)$.

For the first of these terms, using the same estimate as in Appendix 1, we find
\begin{eqnarray*}
&& \frac{N_x}{Z_1\epsilon}\int_{k_1^\prime\geq 0}
\left[\frac{(k_1^{\prime 2}+\kappa^2)^{1/2}-k_1^\prime}{2m}\right]
\frac{k_1^{\prime 2}}{2m}\exp\left\{-\frac{\beta k_1^{\prime 2}}{2m}-\zeta^i
k_1^\prime \right\}dk_1^\prime \\
&\leq& \frac{N_x}{4m^2Z_1\epsilon}\left[\kappa\int_0^{\kappa}
k_1^{\prime 2}\exp\left\{-\frac{\beta k_1^{\prime 2}}{2m}-\zeta^i
k_1^\prime \right\}dk_1^\prime \right.\\
&& + \left. \kappa^2\int_{\kappa}^{\infty}
k_1^\prime \exp\left\{-\frac{\beta k_1^{\prime 2}}{2m}-\zeta^i
k_1^\prime \right\}dk_1^\prime\right] \\
&\leq& \frac{N_x}{4m^2Z_1\epsilon}
\left[\kappa^4 + \kappa^2 M_1(\zeta^1)\right] \\
&=& \frac{N_x}{Z_1\epsilon}\left[(\ell\partial\Phi)^2 + \frac{\ell\partial\Phi}
{2\beta}\right],
\end{eqnarray*}
making it of a smaller order than the last three. Therefore,
$B_3$ itself is of order $O(\ell\log\ell)$.

\subsection{Appendix 4}

The integral in (\ref{energyC}) is bounded by
\begin{eqnarray*}
&& \frac{N_x\kappa^2}{2mZ\epsilon^3}\int_{k_1^\prime\geq 0}
A({\bf k^\prime})d^3{\bf k^\prime} \\
&=& \frac{N_x\kappa^2}{2mZ\epsilon^3}\int_{k_1^\prime\geq 0}
\left(\frac{{\bf k^\prime\cdot k^\prime} }{2m}+\Phi(x^\prime)\right)
\exp\left\{-\frac{\beta{\bf k^\prime\cdot k^\prime}}{2m}
-\mbox{\boldmath$\zeta$}\cdot{\bf k^\prime}\right\}d^3{\bf k^\prime} \\
&=& \frac{N_x\kappa^2}{2mZ\epsilon^3}\left[\frac{(Z_2\epsilon)(Z_3\epsilon)}
{2m}M_2(\zeta^1)+\frac{(Z_1\epsilon)(Z_3\epsilon)}{2m}[M_2(\zeta^2)+
M_2(-\zeta^2)]\right. \\
&& \left. + \frac{(Z_1\epsilon)(Z_2\epsilon)}{2m}[M_2(\zeta^3)+
M_2(-\zeta^3)]+\Phi(x^\prime)(Z_2\epsilon)(Z_3\epsilon)M_0(\zeta^1)\right]. \\
\end{eqnarray*}

Now, to zeroth order in $\zeta$, this reduces to
\[\frac{N_x\kappa^2\pi^{1/2}}{4\sqrt{2}\beta^{3/2}m^{1/2}(Z_1\epsilon)}
+\frac{N_x\kappa^2\pi^{1/2}}{2\sqrt{2}\beta^{3/2}m^{1/2}(Z_2\epsilon)}
 + \frac{N_x\kappa^2\pi^{1/2}}{2\sqrt{2}\beta^{3/2}m^{1/2}(Z_3\epsilon)}
+\frac{N_x\kappa^2}{Z_1\epsilon 2m}\left(\frac{\pi m}{2\beta}\right)^{1/2}.\]

Recalling that $\kappa^2=2\ell m\partial\Phi$ and that the
term $N_x/(Z_i\epsilon)$ is $O(1)$, we find that the expression above
is $O(\ell^2)$.

\subsection{Appendix 5}

As in Appendix 2, the difference between the cases with or without
an external potential is
\begin{equation*}
\frac{k}{m}-\frac{k-(k^2+\kappa^2)^{1/2}}{2m} =
\frac{k+(k^2+\kappa^2)^{1/2}}{2m},
\end{equation*}
so that the integral to be bound in this is
\begin{eqnarray*}
B_5 &:=& \frac{N_x}{Z\epsilon^3}\int_{k_1\leq 0}
\frac{(k_1^2+\kappa^2)^{1/2}+k_1}{2m} \\
&& \times \left(\frac{{\bf k\cdot k} }{2m}
+\Phi(x)\right)
\exp\left\{-\frac{\beta{\bf k\cdot k}}{2m}
-\mbox{\boldmath$\zeta$}\cdot{\bf k}\right\}d^3{\bf k}
\end{eqnarray*}

So performing the change of variable ${\bf k^\prime}=-{\bf k}$,
the integral becomes
\begin{eqnarray*}
B_5 &=& \frac{N_x}{Z\epsilon^3}\int_{k_1^\prime\geq 0}
\frac{(k_1^{\prime 2}+\kappa^2)^{1/2}-k_1^\prime}{2m} \\
&& \times \left(\frac{{\bf k^\prime\cdot k^\prime} }{2m}
+\Phi(x)\right)
\exp\left\{-\frac{\beta{\bf k^\prime\cdot k^\prime}}{2m}
+\mbox{\boldmath$\zeta$}\cdot{\bf k^\prime}\right\}d^3{\bf k^\prime}
\end{eqnarray*}
which reduces to the case dealt in Appendix 3, since the quadratic term in
the exponential quickly dominates the linear one.

\subsection{Appendix 6}

Replacing sums by integrals and using that $p_x(k)=N_x\bar{p}_x(k)$ we find
that the last two terms in (\ref{jMchange}) can be expressed as
\begin{eqnarray*}
&&\partial_j\Phi(x)\frac{N_{x+\ell e_j}}{Z_j(x+\ell e_j)\epsilon}\int_{k\leq 0}
\exp\left\{-\frac{\beta_{x+\ell e_j} k^2}{2m}-\zeta_{x+\ell e_j}^j
k \right\}dk \\
&& + \partial_j\Phi(x-\ell e_j)\frac{N_{x-\ell e_j}}{Z_j(x-\ell_j)\epsilon}
\int_{k\geq\kappa}
\exp\left\{-\frac{\beta_{x-\ell e_j} k^2}{2m}-\zeta_{x-\ell e_j}^j k \right\}dk
\end{eqnarray*}

We first notice that we can move the lower limit of integration in the second
term above from $\kappa$ to zero, since the error involved in doing so is
of order $\kappa$, that is, $O(\ell^{3/2})$. We are then left with
\begin{equation*}
\partial_j\Phi(x)\frac{N_{x+\ell e_j}}{Z_j(x+\ell e_j)\epsilon}M_0
(-\zeta_{x+\ell e_j}^j) + \partial_j\Phi(x-\ell e_j)\frac{N_{x-\ell e_j}}
{Z_j(x-\ell_j)\epsilon}M_0(\zeta_{x-\ell e_j}^j).
\end{equation*}

Expanding $M_0$ to zeroth order in $\zeta$
\begin{equation*}
\partial_j\Phi(x)\frac{N_{x+\ell e_j}}{Z_j(x+\ell e_j)\epsilon}
\left(\frac{\pi m}{2\beta_{x+\ell e_j}}\right)^{1/2}+
\partial_j\Phi(x-\ell e_j)\frac{N_{x-\ell e_j}}{Z_j(x-\ell_j)\epsilon}
\left(\frac{\pi m}{2\beta_{x-\ell e_j}}\right)^{1/2}.
\end{equation*}
In the limit $\ell \rightarrow 0$ this reduces to
\begin{equation*}
\partial_j\Phi(x)\frac{N_x}{Z_j\epsilon}2\left(\frac{\pi m}
{2\beta}\right)^{1/2}=N_x\partial_j\Phi(x)
\end{equation*}

\subsection{Appendix 7}

Here the difference is
\[\frac{k^\prime+(k^{\prime 2}+\kappa^2)^{1/2}}{2m}-\frac{k^\prime}{m}=
\frac{(k^{\prime 2}+\kappa^2)^{1/2}-k^\prime}{2m}
\]
and the integral to bound is
\[B_7:= \frac{N_x}{Z_j\epsilon}\int_{k^\prime}\frac{(k^{\prime 2}
+\kappa^2)^{1/2}-k^\prime}{2m}k^\prime
\exp\left\{-\frac{\beta k^{\prime 2}}{2m}-\zeta^i k^\prime \right\}dk^\prime.\]

Using the same estimate as in Appendix 1, we find
\begin{eqnarray*}
B_7 &\leq& \frac{N_x}{2m Z_j\epsilon}
\left(\kappa \int_0^\kappa k^\prime \exp\left\{
-\frac{\beta k^{\prime 2}}{2m}-\zeta^i k^\prime \right\} dk^\prime \right.\\
&& \left. +\kappa^2\int_\kappa^\infty
\exp\left\{-\frac{\beta k^{\prime 2}}{2m}-\zeta_i k^\prime\right\}dk^\prime
\right)\\
&\leq& \frac{N_x}{2m Z_i\epsilon}[\kappa^3
+\kappa^2 M_0(\zeta^1)],
\end{eqnarray*}
which we conclude is $O(\ell^2)$, since, to
zeroth order in $\zeta$, $M_0$ is proportional to $m^{1/2}$.

\subsection{Appendix 8}

The integral to be bound here is
\[B_8:=\frac{N_x}{Z_j\epsilon}\int_{k\leq 0}\frac{k+(k^2+\kappa^2)^{1/2}}{2m}
k\exp\left\{-\frac{\beta k^2}{2m}-\zeta^i k\right\}dk.\]
With the change of variables $k^\prime=-k$ this becomes
\[B_8 = -\frac{N_x}{Z_j\epsilon}\int_{k^\prime\geq 0}
\frac{k^\prime+(k^{\prime 2}+\kappa^2)^{1/2}}{2m}
k\exp\left\{-\frac{\beta k^{\prime 2}}{2m}-\zeta^i k^\prime \right\}
dk^\prime,\]
which has the same bound as that of Appendix 7.

\end{document}